\documentclass[%
aps,
prx,
reprint,
superscriptaddress,
nofootinbib,
amsmath,
amssymb
]{revtex4-2}
\usepackage{
    physics,
    graphicx,
    multirow,
    tabularx,
    mathtools, 
    placeins, 
    nicefrac, 
    hyperref, 
    threeparttable, 
    siunitx,
    enumitem,
    makecell,
    amsmath,amssymb,amsthm,amsfonts,dsfont
}
\usepackage{mathrsfs}

\usepackage[capitalize]{cleveref}
\Crefname{section}{Sec.}{Secs.}

\usepackage[dvipsnames]{xcolor}
\hypersetup{
    colorlinks,
    linkcolor={blue!50!black},
    citecolor={blue!50!black},
    urlcolor={blue!80!black}
}

\usepackage[caption=false]{subfig} 
\DeclarePairedDelimiter\pbra{\langle\!\langle}{\rvert}
\DeclarePairedDelimiter\pket{\lvert}{\rangle\!\rangle}
\DeclarePairedDelimiterX\pbraket[2]{\langle\!\langle}{\rangle\!\rangle}{#1 \delimsize\vert #2}

\newtheorem{theorem}{Theorem}

\newtheorem{definition}{Definition}

\begin{document}

\title{Non-Markovian Noise Suppression Simplified through Channel Representation}

\author{Zhenhuan Liu}
\email{qubithuan@gmail.com}
\affiliation{Center for Quantum Information, Institute for Interdisciplinary Information Sciences, Tsinghua University, Beijing 100084, China}
\affiliation{Quantum Research Center, Technology Innovation Institute, Abu Dhabi, UAE}

\author{Yunlong Xiao}
\affiliation{A*STAR Quantum Innovation Centre (Q.InC), Agency for Science, Technology and Research (A*STAR), 2 Fusionopolis Way, Innovis \#08-03, Singapore, 138634, Republic of Singapore}
\affiliation{Institute of High Performance Computing (IHPC), Agency for Science, Technology and Research (A*STAR), 1 Fusionopolis Way, Connexis \#16-16, Singapore, 138632, Republic of Singapore}

\author{Zhenyu Cai}
\email{cai.zhenyu.physics@gmail.com}
\affiliation{Department of Engineering Science, University of Oxford, Parks Road, Oxford OX1 3PJ, United Kingdom}
\affiliation{Department of Computing, Imperial College London,
180 Queen’s Gate, London SW7 2AZ, United Kingdom}
\affiliation{Quantum Motion, 9 Sterling Way, London N7 9HJ, United Kingdom}

\begin{abstract}
Non-Markovian noise, arising from memory effects in the environment, poses substantial challenges to conventional quantum noise suppression protocols, including quantum error correction and mitigation.
We introduce a channel representation for arbitrary non-Markovian quantum dynamics, termed the Choi channel, which translates the complex dynamics of non-Markovian noise into the familiar picture of noise channels.
It therefore provides a systematic way to design non-Markovian noise suppression protocols: one can apply existing channel-level error-suppression techniques in the Choi-channel picture and then translate them back to the circuit picture.
The performance of the resulting non-Markovian protocols, including their noise-suppression effects and complexity, can often be inherited directly from the corresponding Choi-channel protocols without requiring a separate analysis.
With this framework, we devise new protocols using Pauli twirling, probabilistic error cancellation, and virtual channel purification.
Pauli twirling reduces non-Markovian noise to noise with only classical temporal correlations; probabilistic error cancellation can fully cancel non-Markovian noise; and virtual channel purification can suppress non-Markovian noise without detailed knowledge of its specific form.
Through these examples, the Choi channel serves as a foundational bridge for systematically converting existing channel-level techniques into non-Markovian noise suppression protocols.

\end{abstract}

\maketitle
\section{Introduction}
Noise presents the most significant obstacle to the practical implementation of quantum computation. A range of techniques has been proposed for noise suppression at different levels, including dynamical decoupling~\cite{violaDynamicalDecouplingOpen1999}, quantum error correction (QEC)~\cite{terhalQuantumErrorCorrection2015,Devitt2013QEC} and mitigation (QEM)~\cite{caiQuantumErrorMitigation2023}. 
Most error suppression techniques are developed based on the assumption that the noise processes occurring at different time slices are independent and can be described as independent quantum channels.
Under this assumption, one can develop QEC and QEM techniques to address the noise channels at a given time slice and simply repeat the QEC and QEM procedure. 
However, since in practice noise is caused by continuous interaction between the environment and the quantum computer, it is no surprise that the environment can exhibit memory effects that lead to classically or even quantumly correlated noise at different time slices~\cite{Rivas2014nonmarkovianity,Shrikant2023nonmarkovianity,Taranto2024characterising}, which are often referred to as non-Markovian noise. Other than certain specific circumstances~\cite{bombin2016resilience,ziyad2025emergent}, in general non-Markovian noise can significantly affect the performance of QEC and QEM protocols~\cite{FKam2025detrimental,kobayashi2024tensor}.

 \begin{figure}
     \centering
     \includegraphics[width=1.0\linewidth]{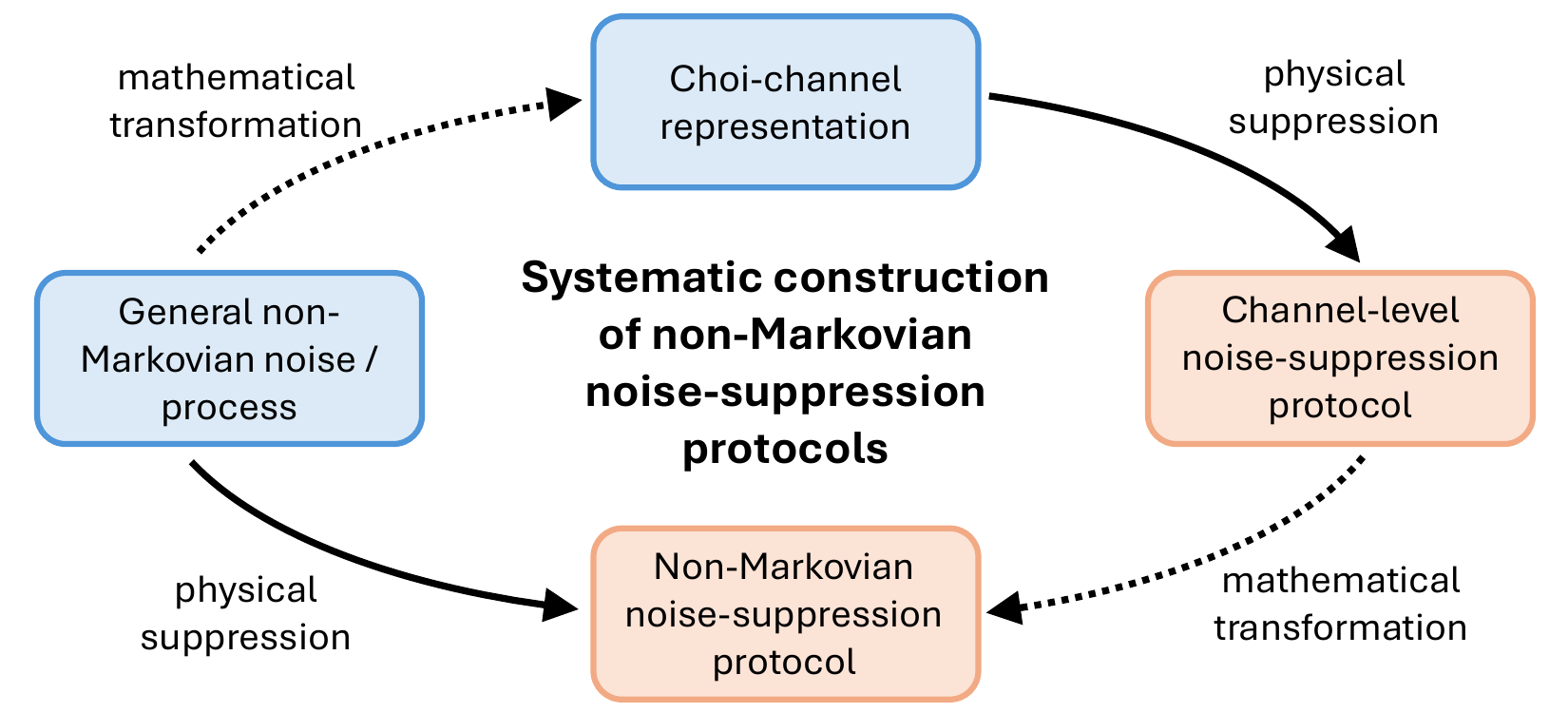}
     \caption{\textbf{Constructing non-Markovian noise suppression protocols from channel noise suppression protocols.}
A general non-Markovian noise process is first mapped, through a purely mathematical transformation, to its Choi-channel representation. 
In this channel picture, one can apply existing channel-level noise suppression protocols. 
The resulting construction is then translated back, again mathematically, to the original circuit picture, yielding an explicit non-Markovian noise suppression protocol. 
The solid arrows denote physical noise suppression steps, while the dotted arrows denote mathematical transformations between the non-Markovian process and the Choi-channel picture.}
     \label{fig:overview}
 \end{figure}

Attempts have been made to deal such noise in several noise-suppression settings, including hardware-level techniques~\cite{violaDynamicalDecouplingOpen1999,terhal2005localnonM,Biercuk2009dynamical,Berk2023dynamicaldecoupling}, QEC~\cite{preskillSufficientConditionNoise2013,tanggara2024strategiccodeunifiedspatiotemporal}, and QEM~\cite{hakoshimaRelationshipCostsQuantum2021}. 
However, these studies typically either rely on specific assumptions about the system-environment interaction or treat non-Markovian effects at an abstract theoretical level without yielding specific protocols.
The primary challenge in analysing general non-Markovian noise lies in its complex nature, which requires higher-order quantum operations for a complete description. These include frameworks such as quantum strategies~\cite{10.1145/1250790.1250873}, quantum combs~\cite{chiribella2008circuit,PhysRevA.80.022339}, process tensors~\cite{PhysRevA.97.012127,PRXQuantum.2.030201}, or quantum circuit fragments~\cite{xing2023fundamentallimitationscommunicationquantum,PhysRevLett.130.240201}. In contrast, most existing error suppression techniques are designed for noise modelled as quantum channels~\cite{nielsen_chuang_2010,wilde_2013,watrous_2018}, making them unsuitable for direct application to non-Markovian scenarios~\cite{PhysRevLett.123.110501,PhysRevResearch.3.023077,PhysRevLett.127.060501,raza2024onlinelearningquantumprocesses}.

In this work, we present a simple yet surprisingly general and effective solution: a mapping between non-Markovian quantum dynamics and quantum channels, which we call the \emph{Choi channel} representation. 
While the Choi \emph{channel} can in principle be further converted to the standard Choi \emph{state} representation of a non-Markovian process~\cite{chiribella2008circuit}, there is no reason to do that since our channel representation is much more natural for studying noise-suppression protocols, given that these protocols are usually constructed to act directly on noise channels, not the Choi state representation of them. 
By applying existing channel-level error-suppression protocols directly in the Choi-channel picture, we can then translate the resulting protocols back to the non-Markovian circuit picture to obtain new protocols for tackling non-Markovian noise, which is shown in Fig.~\ref{fig:overview} and demonstrated through several examples later in this paper.
Moreover, when a channel-level protocol can be translated back to the circuit picture, the error suppression performance and complexity of the resulting non-Markovian noise-suppression protocol is inherited directly from that of the corresponding channel-level protocol, without requiring a separate analysis. 
We begin by developing the $\chi$-matrix representation and Pauli twirling protocol for non-Markovian noise, showing that Pauli twirling can eliminate the quantum temporal correlation in arbitrary non-Markovian quantum dynamics.
Then, using Choi channels, we manage to construct explicit protocols for performing probabilistic error cancellation and purification-based QEM on general non-Markovian noise~\cite{temmeErrorMitigationShortDepth2017,endoPracticalQuantumError2018,hugginsVirtualDistillationQuantum2021,koczorExponentialErrorSuppression2021,liu2024virtualchannelpurification}. 
All of these examples have demonstrated the simplicity and effectiveness of using Choi channels to directly import channel noise suppression techniques for tackling non-Markovian noise, thereby opening up a broad spectrum of new possibilities for addressing non-Markovian noises.

\begin{figure}[htbp]
\centering
\includegraphics[width=0.99\linewidth]{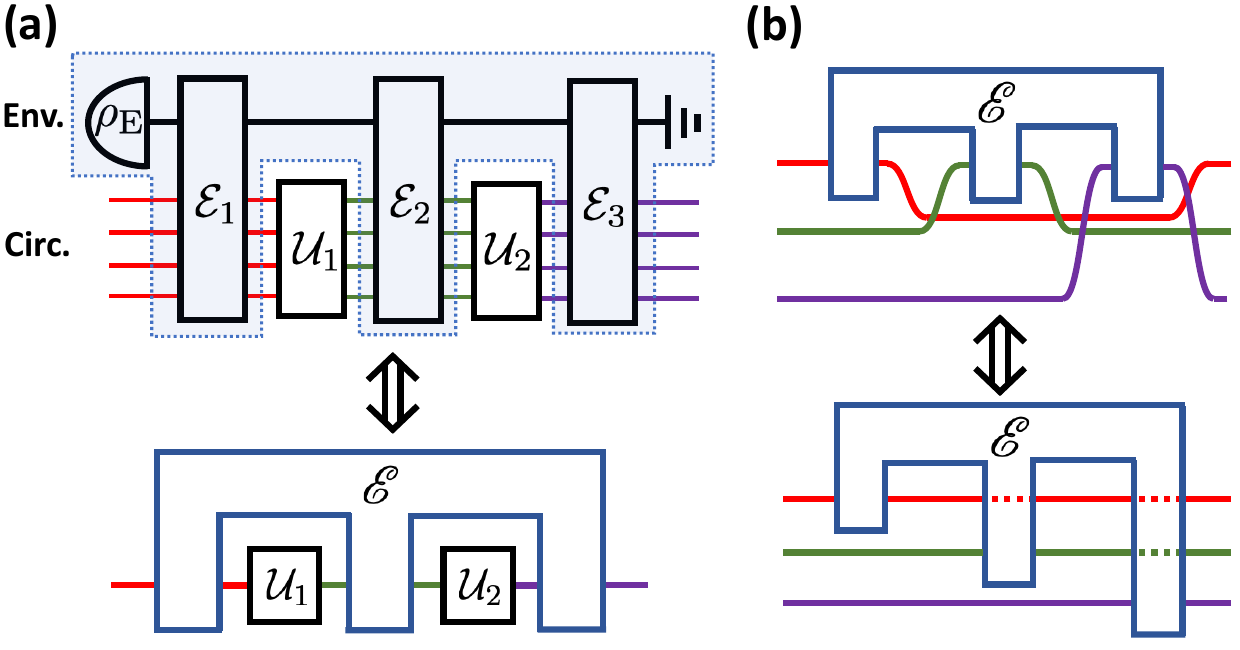}
\caption{(a) Non-Markovian noise caused by interaction with the same environment system at different time steps, where wires in different colours represent systems from different time steps. (b) Shifting the input and output wires of the same time step to the same register outputs the Choi channel. A simplified graph representation is shown in the lower panel. The figure here only shows two circuit layers, which can easily be generalised into an arbitrary number of circuit layers. 
}
\label{fig:choi_channel}
\end{figure}

\section{Choi channel}
No quantum system operates in complete isolation. In a quantum computer, repeated interactions with the environment generate temporal quantum correlations, giving rise to non-Markovian noise that poses significant challenges to reliable information processing. The dynamics of such noise are inherently complex. In general, a single time slice of such noise cannot be adequately described by a quantum channel or even by more general linear maps~\cite{PhysRevLett.73.1060,PhysRevLett.75.3020,PhysRevLett.75.3021}.
We can model a noisy circuit using \cref{fig:choi_channel}(a) where layers of gates, denoted as $\mathcal{U}_i$, are interlaced by layers of noisy interactions, denoted as $\mathcal{E}_i$. In general, these interactions $\mathcal{E}_i$ can be connected through the environment and thus can be quantumly correlated over time. 
We will denote the entire non-Markovian quantum dynamics as $\mathscr{E}$, which can be mapped to a quantum channel  $\mathcal{C}_{\mathscr{E}}$ by shifting the input and output wires of the $i$-th time step to the $i$-th auxiliary register as shown in \cref{fig:choi_channel}(b). 
We will call this the \emph{Choi channel} of the non-Markovian quantum dynamics.

\begin{figure}[htbp!]
\centering
\includegraphics[width=1\linewidth]{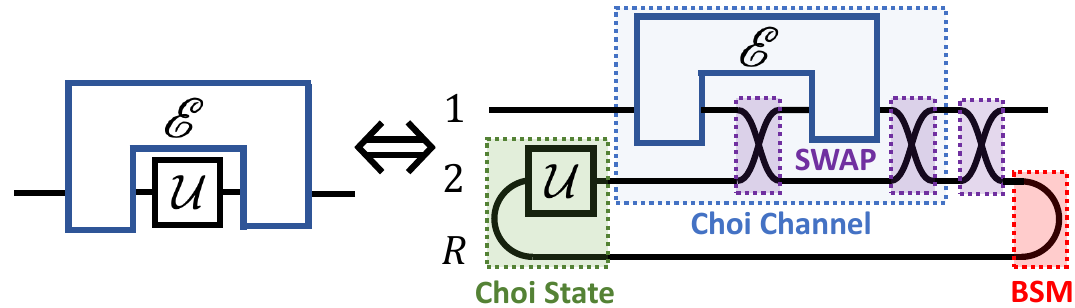}
\caption{The mapping between non-Markovian quantum dynamics and its Choi channel. BSM stands for Bell state measurement. }
\label{fig:choi_iso}
\end{figure}

For simplicity, let us illustrate this using the example of a non-Markovian noise process $\mathscr{E}$ interacting at two different time steps, before and after the ideal gate layer $\mathcal{U}$, as shown in \cref{fig:choi_iso}. 
Such non-Markovian noise affecting two points in time can be mathematically represented by a \emph{superchannel} that maps one channel to another~\cite{Chiribella_2008}.
By applying the Choi channel of the noise $\mathcal{C}_{\mathscr{E}}$ to the Choi state of the ideal gate layer $\rho_{\mathcal{U}}$ and performing the relevant Bell state measurements, we can obtain the output channel representing the noisy gate layer. 
This approach closely mirrors the action of a conventional noise channel on some input state. Our example here can be more formally stated as follows.
\begin{definition}\label{def:choi_channel}
Given a superchannel $\mathscr{E}$, its Choi channel is defined as 
\begin{equation}\label{eq:choi_def}
\mathcal{C}_{\mathscr{E}} = \mathrm{SWAP}^{(1,2)}\circ(\mathscr{E}^{(1)}\otimes\mathscr{I}^{(2)})[\mathrm{SWAP}^{(1,2)}],
\end{equation}
where $\mathscr{I}$ is the identity superchannel, $1$ and $2$ denote two quantum registers, and $\circ$ denotes the composition between channels.
The resultant channel of acting $\mathscr{E}$ on the input channel $\mathcal{U}$ is
\begin{equation}\label{eq:choi_to_super}
\begin{split}
    \mathscr{E}[\mathcal{U}](\rho)=\Tr_{2,R}&\Big\{(\mathbb{I}^{(1)}\otimes \Phi^{(2,R)}_+)\big[\big(\mathrm{SWAP}^{(1,2)} \\
&\quad \circ \mathcal{C}_{\mathscr{E}}^{(1,2)}\big)\otimes\mathcal{I}^{(R)}\big](\rho^{(1)} \otimes \rho_{\mathcal{U}}^{(2,R)})\Big\},
\end{split}
\end{equation}
where $\ket{\Phi_+}=\sum_i\ket{ii}$ is the unnormalized maximally entangled state, $\mathcal{I}^{(R)}$ represents the identity channel acting on the reference system $R$, $\mathbb{I}$ is the identity matrix, and $\rho_{\mathcal{U}}=\mathcal{U}\otimes\mathcal{I}(\Phi_+)$ is the unnormalized Choi state of $\mathcal{U}$.
\end{definition}

While our demonstration focuses on superchannels, representing the sequential composition of two successive quantum processes, our framework is broadly applicable and readily extends to address non-Markovian noise arising from the composition of arbitrary finite quantum processes. It provides a universal characterization of quantum noise models that respect a causal structure. Markovian noise, temporally correlated noise at specific time steps, and other similar phenomena are naturally included as special cases within this framework. Additional details, including the role of the extra SWAP in \cref{eq:choi_to_super}, are presented in Methods. Note that, although the Choi channel represents a valid physical channel, it is not arbitrary. Its dynamics is fundamentally constrained by the non-signalling principle, which prohibits the transmission of information from the input or measurement in the later register to the quantum process in the earlier register.

Although Eq.~\eqref{eq:choi_def} is primarily a mathematical equivalence, it also gives an efficient way to realize the Choi channel by applying the superchannel to a SWAP channel and then rearranging the output registers.
This procedure contains no mechanism that would increase the sample complexity, such as additional quantum measurements or sampling over extra random variables.
This observation is useful for analyzing the performance of the non-Markovian noise-suppression protocols constructed in this work, including the sample complexity and noise suppression rate. 
Whenever a channel-level protocol acting on the Choi channel can be translated back to a causal circuit-level protocol for the original non-Markovian noise, the output statistics of the two descriptions are identical. 
Consequently, the sample complexity and noise suppression rate of the resulting non-Markovian protocol is the same as that of the corresponding channel-level protocol in the Choi-channel picture. 
This equivalence of complexity is illustrated explicitly in the examples below, including Pauli twirling, probabilistic error cancellation, and virtual channel purification.

\section{Twirling non-Markovian noise }
Pauli twirling is a standard technique in noise suppression for converting coherent errors into stochastic Pauli errors~\cite{wallmanNoiseTailoringScalable2016,caiConstructingSmallerPauli2019,tsubouchi2024symmetric}. 
Operationally, in each circuit run, one samples a Pauli operator $G_m$ uniformly at random from the relevant Pauli group and inserts the sampled Pauli gate before and after the target noise process. 
Since Pauli operators are self-inverse, they act trivially when noiseless. With noise, inserting the same Pauli on both sides realizes the Pauli-conjugated noise process $\mathcal{G}_m \circ \mathcal{E} \circ \mathcal{G}_m$ for that run. 
The randomness is entirely classical: each shot simply uses one sampled Pauli configuration, and the twirled channel corresponds to the ensemble average over this classical randomization.

\begin{figure}[htbp]
\centering
\includegraphics[width=0.95\linewidth]{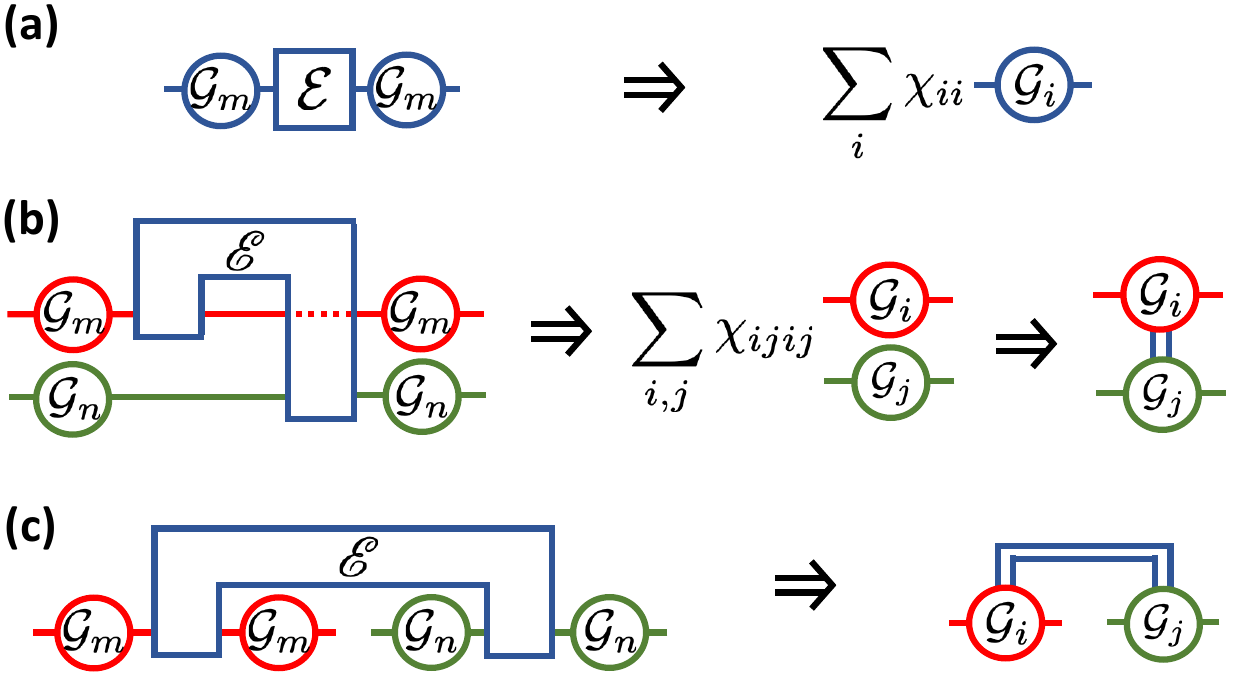}
\caption{(a) Pauli twirling of noise channel. (b) Pauli twirling of Choi channel. (c) Pauli twirling of non-Markovian noise, obtained by rearranging the wires in the Choi-channel picture in (b) back to the circuit picture. The blue double solid lines denote the classical correlations.}
\label{fig:pauli_twirling}
\end{figure}

Any noise channel $\mathcal{E}$ can be written in the $\chi$-matrix representation
\begin{align}
    \mathcal{E}(\cdot)
    =
    \sum_{i,j}\chi_{ij}G_i(\cdot)G_j ,
\end{align}
where $G_i$ and $G_j$ are Pauli operators and $\chi_{ij}$ is a positive semi-definite matrix with $\sum_i \chi_{ii}=1$. 
When taking the averaging over all possible Pauli operators $G_m$ for twirling, we have
\begin{align}\label{eqn:Pauli_twirl}
    \mathsf{T}\left[\mathcal{E}\right]
    =
    \mathbb{E}_{m}
    \left(
        \mathcal{G}_m \circ \mathcal{E} \circ \mathcal{G}_m
    \right)
    =
    \sum_i \chi_{ii}\mathcal{G}_i ,
\end{align}
where $\mathcal{G}_i$ denotes the channel corresponding to the Pauli operator $G_i$. 
Thus, Pauli twirling removes the off-diagonal entries of the $\chi$ matrix and converts the noise channel into a classical mixture of Pauli errors, as illustrated in Fig.~\ref{fig:pauli_twirling}(a).

The overhead of Pauli twirling is only a gate-level overhead, not a sample-complexity overhead. 
Pauli operators are tensor products of single-qubit Pauli operators, so the inserted twirling gates do not require additional entangling operations and can often be implemented with negligible additional noise or absorbed into the original quantum gate~\cite{wallmanNoiseTailoringScalable2016}. 
The reason that no additional sample complexity is introduced is that the Pauli label is sampled as part of the same experimental shot. 
For each shot, one samples the Pauli gates, runs the Pauli-dressed circuit, and records the same measurement outcome as in the original experiment. 
Thus the outcome range is unchanged. 
By standard concentration bounds, estimating the averaged twirled process in Eq.~\eqref{eqn:Pauli_twirl} therefore requires the same shot scaling as the original measurement task, with no extra factor depending on the system dimension, the size of the Pauli group, or the number of twirling locations.

Turning to the non-Markovian case, we use superchannel $\mathscr{E}$, which models the quantum dynamics as two interconnected processes mediated by a quantum memory, as an illustrative example. 
Its Choi channel can be written in the $\chi$-matrix representation just like any other channel
\begin{align}\label{eqn:choi_chi}
   \mathcal{C}_{\mathscr{E}}(\cdot) = \sum_{i,j,k,\ell}\chi_{ijk\ell}(G_i\otimes G_{j})(\cdot)(G_k\otimes G_{\ell})
\end{align}
where we have written the Pauli basis in terms of the tensor product of Paulis on the two registers of the Choi channel. 
As shown in \cref{fig:pauli_twirling}(b), we can Pauli twirl this Choi channel by conjugating it with random Pauli gates just like noise channels, which makes the Choi channel Pauli-diagonal, $\mathsf{T} \left[\mathcal{C}_{\mathscr{E}}\right] =  \sum_{i,j} \chi_{ijij} \mathcal{G}_i \otimes \mathcal{G}_j$.
We can translate this twirling process back into the circuit picture as shown in \cref{fig:pauli_twirling}(c), which can be more formally stated as the following.
\begin{theorem}
We can represent any quantum superchannel $\mathscr{E}$ using $\chi$-matrix representation
\begin{equation}\label{eq:chi}
\mathscr{E}[\mathcal{U}](\rho)=\sum_{i,j,k,\ell}\chi_{ijk\ell}G_j\mathcal{U}(G_i \rho G_k)G_\ell.
\end{equation}
where $\mathcal{U}$ and $\rho$ are some arbitrary input channel and state, respectively. The way we perform Pauli twirling on this superchannel and the resultant superchannel is given by:
\begin{equation}\label{eq:def_pauli_twirling}
\begin{split}
\mathsf{T}\{\mathscr{E}\}[\mathcal{U}]&= \mathbb{E}_{m, n} \left\{\mathcal{G}_{n}\circ\mathscr{E}[\mathcal{G}_{n}\circ\mathcal{U}\circ\mathcal{G}_{m}] \circ \mathcal{G}_{m}\right\}\\
& =\sum_{i,j}\chi_{ijij}\mathcal{G}_j\circ\mathcal{U}\circ\mathcal{G}_i.
\end{split}
\end{equation}
\end{theorem}
Hence, performing Pauli twirling on non-Markovian noise is \emph{the same as performing Pauli twirling on individual noise channels.} 
Specifically, as shown in Fig.~\ref{fig:pauli_twirling}(c), each occurrence of the non-Markovian noise is sandwiched by an independently and uniformly sampled pair of Pauli gates. 
The definition in Eq.~\eqref{eq:def_pauli_twirling} has a similar structure as the ordinary Pauli-twirling formula in Eq.~\eqref{eqn:Pauli_twirl}. 
Therefore, the same argument applies: the Pauli labels are sampled within each experimental shot and do not enlarge the measurement outcome range, so the sample complexity is unchanged relative to running the original circuit without twirling. 
The remaining cost is only the gate-level cost of inserting Pauli gates at the different time steps.

Note that $\{\chi_{ijij}\}_{i,j}$ constitute a probability distribution since $\chi_{ij,kl}$ is a positive matrix and $\sum_{i,j}\chi_{ijij}=1$. 
The output channel in \cref{eq:def_pauli_twirling} is thus Pauli noise happening at different time steps with classical correlations. All quantum coherence between the two time steps has been removed. Ref.~\cite{winickConceptsConditionsError2022,figueroa-romeroOperationalMarkovianizationRandomized2024} have also discussed that applying Pauli twirling can remove the quantum coherence across time, but their description of non-Markovian noise is complexified by the needs to include the environment, resulting in a less explicit relation for non-Markovian noise before and after twirling. We are able to show a simplified relationship here using the Choi-channel and $\chi$-matrix representation of the non-Markovian noise. 
Although demonstrated using non-Markovian noises affecting two time steps, all results in this work, including the $\chi$-matrix and Pauli twirling technique, can be directly extended to general non-Markovian quantum dynamics across multiple time steps.

Our result for twirling non-Markovian noise has direct implications for QEC under temporally correlated noise. 
In standard QEC, reliable syndrome extraction usually requires several rounds of syndrome measurements, so temporal correlations between different rounds of syndrome measurement outcomes can strongly affect the decoding performance~\cite{dennisTopologicalQuantumMemory2002}. 
Single-shot QEC provides a different route: by using redundant syndrome information, or meta-checks, it can infer and control measurement error from a single round of syndrome extraction. 
Bomb\'{i}n proved that such single-shot QEC codes exhibit a threshold for quantum memory under stochastic noise that is spatially local while allowing arbitrary classical temporal correlations~\cite{bombin2016resilience}. 
The non-Markovian Pauli-twirling protocol introduced above transforms arbitrary non-Markian noise into stochastic Pauli noise with only classical temporal correlations. 
Therefore, after twirling, the resulting noise falls into the class of temporally correlated stochastic noise covered by Bomb\'{i}n's single-shot QEC threshold result.
In the context of preserving quantum memory, the additional random Pauli gates introduced during twirling can be merged with the gates required for implementing the QEC circuit~\cite{wallmanNoiseTailoringScalable2016} and thus will not introduce much gate overhead and also come with no sampling overhead. 
Therefore, we are able to extend results in Ref.~\cite{bombin2016resilience} to all possible non-Markovian noise, not just classically correlated ones, i.e. \emph{one can correct arbitrary non-Markovian noise by combining non-Markovian noise Pauli twirling with single-shot QEC.}

\section{Mitigating non-Markovian noise}
\begin{figure}[htbp]
\centering
\includegraphics[width=1\linewidth]{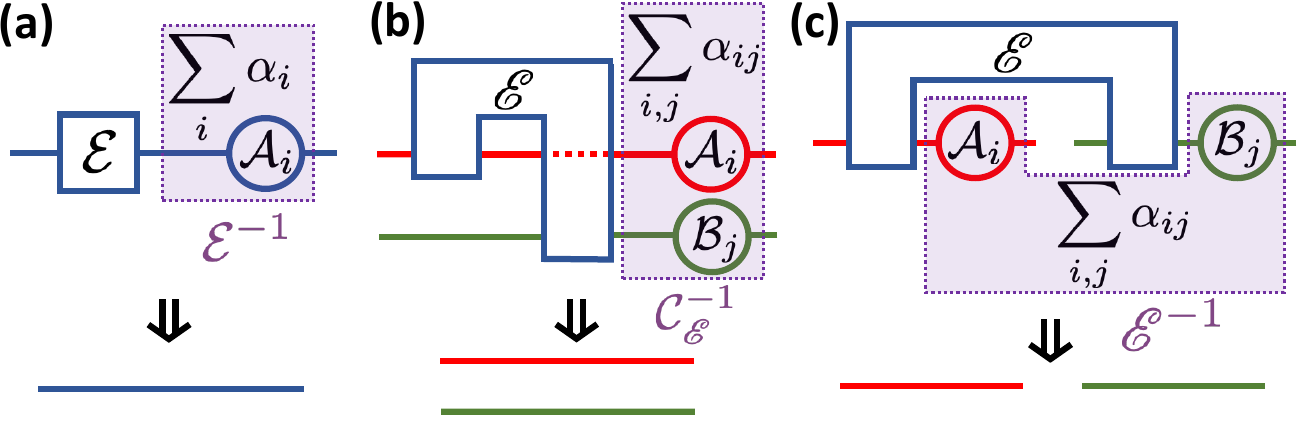}
\caption{
Probabilistic error cancellation for (a) noise channel; (b) Choi channel; (c) non-Markovian noise. 
Panel (c) is obtained from panel (b) by a simple rearrangement of the Choi-channel wires back to the circuit picture. 
Here, the straight lines represent the identity (noiseless) channel.}
\label{fig:pec}
\end{figure}

If we know the exact form of the error channel $\mathcal{E}$, we can try to apply its inverse to mitigate the damage caused by the noise. However, such an inverse channel is often not completely positive and trace-preserving and thus cannot be implemented physically. Probabilistic error cancellation is a way to virtually implement the inverse map $\mathcal{E}^{-1}$ through linear combination of physically implementable operations $\{\mathcal{A}_i\}$ with post-processing~\cite{temmeErrorMitigationShortDepth2017,endoPracticalQuantumError2018}, as shown in \cref{fig:pec}(a). 
The cost of PEC is determined by the quasiprobability coefficients $\alpha_i$ in this linear decomposition: the larger the total absolute weight of these coefficients, or equivalently the larger the negative contribution required in the decomposition, the higher the sampling overhead~\cite{Jiang2021physical}.

Using the Choi channel representation of the non-Markovian noise $\mathscr{E}$, we can straightforwardly obtain its inverse map $\mathcal{C}_{\mathscr{E}}^{-1}$. It can be decomposed into basis operations generated using $Z$-basis measurement, $S$ gate and Hadamard as outlined in \cite{endoPracticalQuantumError2018}, which means that these basis operators can be written as tenor products of local basis in each register, denoted using $\{\mathcal{A}_{i}\}$ and $\{\mathcal{B}_{j}\}$, as shown in Fig.~\ref{fig:pec}(b):
\begin{align}
    \mathcal{C}_{\mathscr{E}}^{-1} = \sum_{ij} \alpha_{ij}\mathcal{A}_{i}\otimes \mathcal{B}_{j}.
\end{align}
With such a tensor representation, we can directly translate this inverse Choi channel back into the circuit picture to implement the inverse superchannel:
\begin{align}
   \mathscr{E}^{-1} \circ \mathscr{E}\left[\,\mathcal{U}\,\right] = \sum_{i, j} \alpha_{ij} \mathcal{B}_{j}\circ\mathscr{E}[\,\mathcal{U}\circ\mathcal{A}_{i}\,]
\end{align}
which is shown in \cref{fig:pec} (c). 

To summarize, the non-Markovian PEC protocol is implemented by sampling a pair of basis operations $(\mathcal A_i,\mathcal B_j)$ according to the quasiprobability coefficients $\alpha_{ij}$. 
For each sampled pair, $\mathcal A_i$ is inserted after the first time step of the non-Markovian noise, and $\mathcal B_j$ is inserted after the second time step, as illustrated in Fig.~\ref{fig:pec}(c). 
Averaging these circuit-picture operations with coefficients $\alpha_{ij}$ virtually implements the inverse superchannel and cancels the original non-Markovian noise. 
Since this construction is exactly the translation of the Choi-channel PEC protocol, the same coefficients $\alpha_{ij}$ determine the quasiprobability overhead in both pictures. 
Therefore, the non-Markovian PEC protocol has the same sampling complexity as the corresponding Choi-channel PEC protocol~\cite{wang2025mitigation}. 
This also reflects the intrinsic difficulty of treating non-Markovian noise compared with ordinary channel noise: for example, inverting a single-qubit non-Markovian noise process involving $t$ time steps is equivalent to inverting some global noise channel acting on $t$ qubits in the Choi-channel picture.

The main limitation of the non-Markovian PEC protocol is that constructing the inverse map requires detailed knowledge of the non-Markovian noise process. 
As the number of time steps increases, temporal correlations make the description of the noise increasingly complicated, with the number of parameters growing exponentially in the number of correlated time steps. 
This substantially increases the cost of noise characterization~\cite{White2020demonstration,white2022ptt,White2025unifying}. 
The Pauli-twirling protocol discussed above can partially alleviate this problem. 
It does not rely on the detailed form of the non-Markovian noise and removes the off-diagonal elements of the $\chi$ matrix, leaving only the Pauli-diagonal components. 
This diagonal structure also makes the inverse map much easier to construct. 
However, the number of remaining diagonal coefficients can still scale exponentially for large systems or many time steps. 
It is therefore desirable to develop non-Markovian noise-suppression protocols that do not require detailed noise characterization.

\section{Purifying non-Markovian noise.}
\begin{figure}[htbp]
\centering
\includegraphics[width=0.99\linewidth]{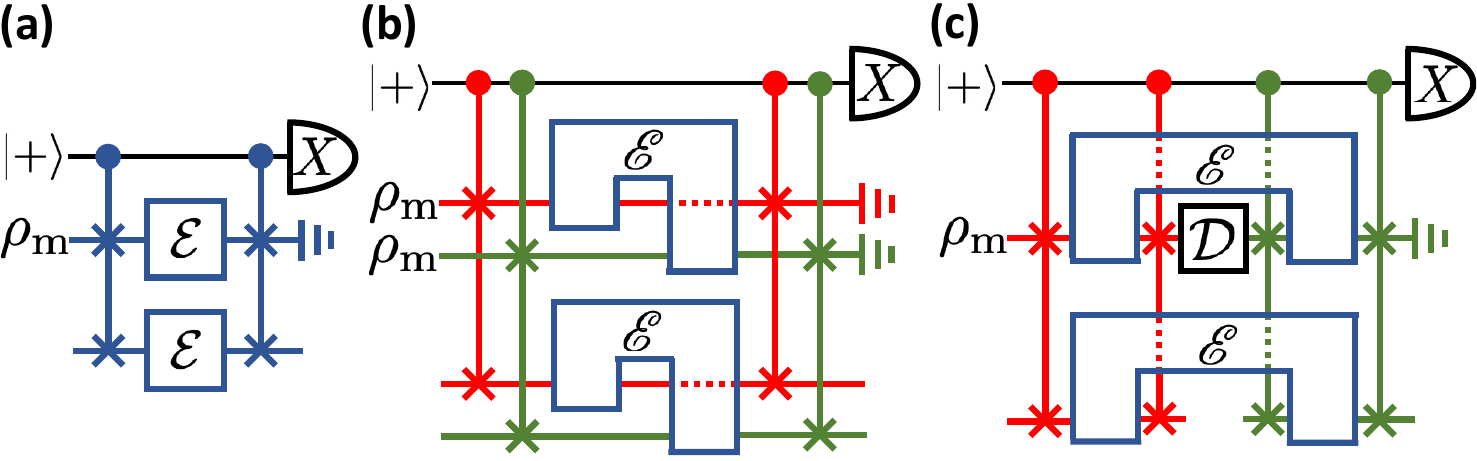}
\caption{
Virtual purification at the level of (a) noise channel; (b) noise Choi channel; (c) non-Markovian noise. 
Panel (c) is obtained from panel (b) by rearranging the Choi-channel wires back to the circuit picture.
Here, $\rho_m$ represents the maximally mixed state and $\mathcal D$ is the completely depolarising channel.}
\label{fig:vcp}
\end{figure}

Virtual channel purification~\cite{liu2024virtualchannelpurification} is a calibration-free channel noise-suppression technique. 
The circuit for implementing virtual channel purification is shown in \cref{fig:vcp}(a), where two copies of the target noise channel are coherently coupled by two CSWAP gates controlled by an ancillary qubit. 
One copy of the noise channel acts on the main register, while the other acts on a maximally mixed input state, which can be prepared by random initialization and is discarded at the end of the circuit. 

Let $p_+$ and $p_-$ be the probabilities of obtaining the $\ket{+}$ and $\ket{-}$ outcomes when measuring the control qubit in the Pauli-$X$ basis, and let $\mathcal E_+$ and $\mathcal E_-$ denote the corresponding resultant channels after measurements on the main register. 
For a Pauli-diagonal noise channel $\mathcal E=\sum_i p_{i}\mathcal G_i$, the post-processed channel is given by
\begin{equation}\label{eq:channel_processing}
    \mathcal{E}_{\mathrm{vp}}=\frac{p_+\mathcal E_+-p_-\mathcal E_-}{p_+-p_-}
    =
    \frac{\sum_i p_i^2 \mathcal G_i}{\sum_j p_j^2}.
\end{equation}
In the small-noise regime, where the noiseless component $p_{0}$ is much larger than the other coefficients, this protocol suppresses the noise rate from first order to second order, i.e., $p_i$ to $p_i^2$. 
The sampling overhead is governed by the normalization factor and scales approximately as $1/\sum_jp_j^2$. 
This protocol has been experimentally implemented~\cite{fei2026vcp}, and the noise-suppression effect can be exponentially enhanced by coherently processing more copies of the noisy channel. 
Importantly, the implementation of the protocol does not require any detailed knowledge of the noise, beyond the standard assumption in quantum noise suppression that the dominant component is the noiseless one. 
This motivates us to extend such a calibration-free purification protocol to non-Markovian noise.

We can directly apply virtual channel purification to the Choi channel of the non-Markovian noise, as shown in \cref{fig:vcp}(b). 
Since the Choi-channel contains multiple parties corresponding to different time steps of the original non-Markovian noise, the number of CSWAP operations also increases, as represented by the green and red lines in \cref{fig:vcp}(b).
When translating this construction back to the circuit picture, these CSWAP operations are placed around the noise operations at different time steps of the non-Markovian process, as shown in Fig.~\ref{fig:vcp}(c). 
Meanwhile, the maximally mixed input on the green registers in the Choi-channel picture becomes a completely depolarising channel $\mathcal D$ in the circuit picture, which can be implemented by applying random Pauli gates without increasing the sampling overhead. 
The post-processing is the same as in the channel-level virtual purification protocol. 
We use Pauli-diagonal non-Markovian noise, which can be obtained by Pauli twirling, as an example to illustrate the effect of virtual channel purification.
\begin{theorem}
Given a Pauli-diagonal superchannel
\begin{equation}
    \mathscr{E}[\cdot]
    =
    \sum_{i,j}p_{ij}\,
    \mathcal{G}_i\circ\cdot\circ\mathcal{G}_j ,
\end{equation}
performing the circuit shown in Fig.~\ref{fig:vcp}(c) gives two conditional superchannels on the main register, depending on the Pauli-$X$ measurement outcome of the control qubit. 
Let $p_+$ and $p_-$ be the probabilities of obtaining the $\ket{+}$ and $\ket{-}$ outcomes, respectively, and let $\mathscr{E}_+$ and $\mathscr{E}_-$ denote the corresponding conditional superchannels on the main register. 
By taking the virtual difference between the two measurement branches, one obtains
\begin{equation}\label{eq:virtual_comb_purification}
    \mathscr{E}_{\mathrm{vp}}[\cdot]
    =
    \frac{
    p_+\mathscr{E}_+
    -
    p_-\mathscr{E}_-
    }{
    p_+ - p_-
    }[\cdot]=
    \frac{\sum_{i,j}p_{ij}^2}{\sum_{i^\prime,j^\prime}p_{i^\prime j^\prime}^2}
    \mathcal{G}_i\circ\cdot\circ\mathcal{G}_j.
\end{equation}
\end{theorem}

The same post-processing formula as Eq.~\eqref{eq:channel_processing} appears in Eq.~\eqref{eq:virtual_comb_purification}, with the channel coefficients $p_i$ replaced by the Pauli-diagonal coefficients $p_{ij}$ of the non-Markovian superchannel. 
Therefore, the noise-suppression effect is inherited directly from the channel-level virtual purification protocol: when the noiseless component is dominant, the relative weights of the noisy components are suppressed to second order.

The sampling complexity has the same origin as in virtual channel purification. 
In both cases, the virtual estimator is obtained by taking the difference between the two control-measurement branches and normalizing it by the factor $p_+-p_-$. 
This denominator amplifies the variance of the estimator and therefore determines the sampling overhead. 
Because the Choi-channel representation is mathematically equivalent to the original non-Markovian process, the values of $p_+$ and $p_-$ computed in the Choi-channel picture are the same as those obtained after translating the protocol back to the non-Markovian circuit picture. 
Thus, the variance amplification and sampling complexity are also the same as in the corresponding Choi-channel protocol, with an overhead of the same form as virtual channel purification, scaling approximately as
\begin{equation}
    \frac{1}{\sum_{i,j}p_{ij}^{2}} .
\end{equation}
A more detailed complexity analysis, together with the experimental implementation of this protocol, can be found in Ref.~\cite{liu2026vcombp}.

\section{Discussion}
In this work, we propose a systematic approach for constructing non-Markovian noise suppression protocols based on the Choi channel -- a channel-level isomorphism that faithfully represents arbitrary non-Markovian quantum dynamics. Since most existing error suppression methods are designed specifically for quantum channels, this isomorphism enables the direct extension of many established techniques to the non-Markovian setting. 
We illustrate this approach with concrete protocols based on Pauli twirling, probabilistic error cancellation, and virtual channel purification.
An important advantage of this approach stems from the isomorphism given in Def.~\ref{def:choi_channel}, which ensures that key performance metrics of the resulting non-Markovian protocols, such as the noise suppression rate and sample complexity, are equivalent to those of their Choi-channel counterparts. 
Therefore, to theoretically guarantee the performance of a non-Markovian suppression protocol, one may begin with a channel-level protocol whose performance has already been rigorously established.

This perspective is analogous to the role of different representations of quantum channels. 
A quantum channel can be equivalently described by Kraus operators, a Choi matrix, a Liouville superoperator, or a Stinespring dilation~\cite{woodTensorNetworksGraphical2015}. 
Although these descriptions are mathematically equivalent, they highlight different physical aspects: the Choi matrix connects channel properties to state properties, while the dilation picture explains the channel as arising from an interaction with an environment. 
Similarly, non-Markovian processes can be represented by quantum combs~\cite{PhysRevLett.101.060401,PhysRevA.80.022339}, process tensors~\cite{PhysRevA.97.012127,PRXQuantum.2.030201}, Choi states~\cite{chiribella2008circuit}, and, as used here, Choi channels. 
The Choi-channel representation does not merely provide another equivalent description; it gives a channel-based viewpoint that is particularly suited to importing and analysing noise-suppression protocols. 
Developing other representations adapted to different tasks may therefore be a useful direction for studying non-Markovian quantum dynamics beyond noise suppression.

An exciting avenue for future work involves applying the Choi channel formalism to specific QEC protocols to study their behaviour under non-Markovian noise. While this will likely involve complex analysis, the Choi channel framework should still offer a simplified and more general analysis compared to prior approaches, which are often constrained to specific noise models. 
We also anticipate that Choi channel will have broad applications in many other tasks related to non-Markovian quantum dynamics, including quantum algorithms~\cite{zhu2024reversing,yoshida2024universal}, open quantum systems~\cite{davies1974markovian,breuer2002theory,wang2024markovian}, randomized benchmarking~\cite{helsen2022RB,figueroa2021nonmarkovianRB,FigueroaRomero2022towardsgeneral} and classical shadow~\cite{huangPredictingManyProperties2020,chenRobustShadowEstimation2021,farias2024robust,hu2024demonstration}. In the process, one can straightforwardly define many other useful representations for non-Markovian noise like the Pauli transfer matrix and Kraus representation though Choi channels, just like what we have done for $\chi$-matrix representation. All of these tools have the potential to greatly simplify the analysis of non-Markovian noise under different contexts. There are numerous connections between our formalism and higher-order quantum operations, as discussed in Methods. It would be highly interesting to explore the potential of incorporating additional results from there into our framework for analyzing non-Markovian noise.

\section{Acknowledgement}
We highly appreciate the insightful discussions with Ingo Roth, Suguru Endo, Yuichiro Matsuzaki, Ziwen Liu, Ryuji Takagi, Xun Gao, Qiushi Liu, Fei Meng, Yuxiang Yang and Hong-Ye Hu.
ZL acknowledges support from the National Natural Science Foundation of China Grant No.~12174216 and the Innovation Program for Quantum Science and Technology Grant No.~2021ZD0300804 and No.~2021ZD0300702. 
YX is supported by the National Research Foundation, Singapore, and A*STAR through the Quantum Engineering Programme (NRF2021-QEP2-02-P03), and by A*STAR through its Central Research Funds and Career Development Fund (C243512002).
ZC acknowledges support from the EPSRC QCS Hub EP/T001062/1, EPSRC projects Robust and Reliable Quantum Computing (RoaRQ, EP/W032635/1), Software Enabling Early Quantum Advantage (SEEQA, EP/Y004655/1) and the Junior Research Fellowship from St John’s College, Oxford.


%

\appendix

\section{Generalisation beyond two time steps}
\begin{figure}[htbp]
\centering
\includegraphics[width=1\linewidth]{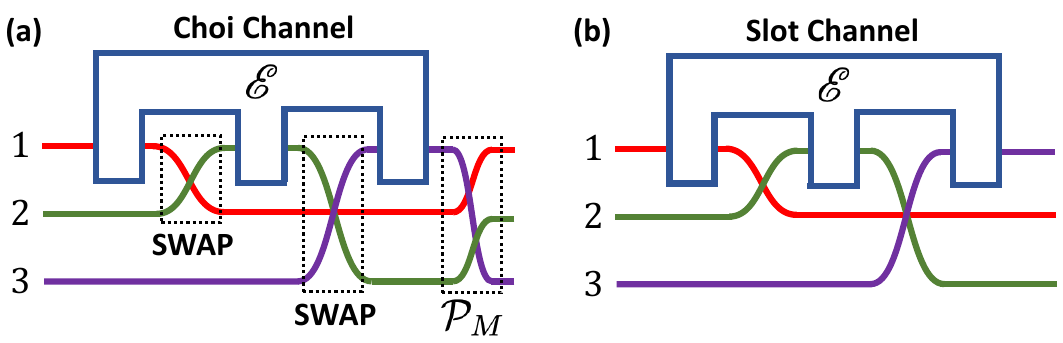}
\caption{Graphical representations for Choi channel and slot channel.}
\label{fig:slot_channel}
\end{figure}

We will generalise many of the results in the main text beyond two time steps. For a given quantum circuit that is affected by some non-Markovian noise as shown in \cref{fig:choi_channel}(a), we can mathematically represent the non-Markovian noise as a $3$-comb $\mathscr{E}$ with its individual tooth being the noise process at the individual error location. Generally, an $M$-comb $\mathscr{E}$ has $M$ teeth and $M-1$ slots. In the main text, we have discussed that such a comb can be mapped to its Choi channel $\mathcal{C}_{\mathscr{E}}$ by shifting the input and output wires of the $m$-th tooth to the $m$-th register. More formally, the Choi channel $\mathcal{C}_{\mathscr{E}}$ of a general $M$-comb $\mathscr{E}$ is given by
\begin{equation}
    \begin{split}
        \mathcal{C}_{\mathscr{E}} = \mathcal{P}_M \circ (\mathscr{E}^{(1)} \otimes \mathscr{I}^{(2,...M)})\big[\mathrm{SWAP}^{(1,2)}, \mathrm{SWAP}^{(1,3)},&\\
    ... , \mathrm{SWAP}^{(1,M)}\big]&
    \end{split}
\end{equation}
where $\mathscr{I}^{(2,...M)}$ is the identity comb that acts on the registers $2$ to $M$, the $m$th channel in the square brackets are input into the $m$-th slot of the comb in front, and $\mathcal{P}_M$ is the cyclic permutation operator that maps $m$ to $m-1$ for $m = 2, ... M$ and maps $1$ to $M$. We have explicitly drawn the diagram for a $3$-comb in \cref{fig:slot_channel}(a).

Such a channel representation with the input and output wire of the same teeth on the same register is great for analysing protocols that act on the (teeth of the) noise like the twirling and QEM protocols discussed in the main text. When trying to obtain the output noisy circuit as a result of applying such a non-Markovian noise to some input noiseless circuit layers, we are trying to fit circuit layers into the \emph{slots}, not the teeth, of the non-Markovian noise. For the purpose of studying the result of applying the non-Markovian noise, it is thus more natural to use another channel representation where we shift the input and output wires of the $m$-th slot (not tooth) to the $(m+1)$-th register will be more natural, and we will simply call it the \emph{slot channel}. The slot channel of the same non-Markovian noise $\mathcal{C}_{\mathscr{E}}$ can be explicitly written as
\begin{equation}
    \begin{split}
        \mathcal{C}_{\mathscr{E}}^{S} &= (\mathscr{E}^{(1)} \otimes \mathscr{I}^{(2,...M)})\big[\mathrm{SWAP}^{(1,2)}, 
    ... , \mathrm{SWAP}^{(1,M)}\big] \\
    &= \mathcal{P}_M^{-1} \circ \mathcal{C}_{\mathscr{E}},
    \end{split}
\end{equation}
which is graphically shown in \cref{fig:slot_channel}(b).

\begin{figure}[htbp]
\centering
\includegraphics[width=1\linewidth]{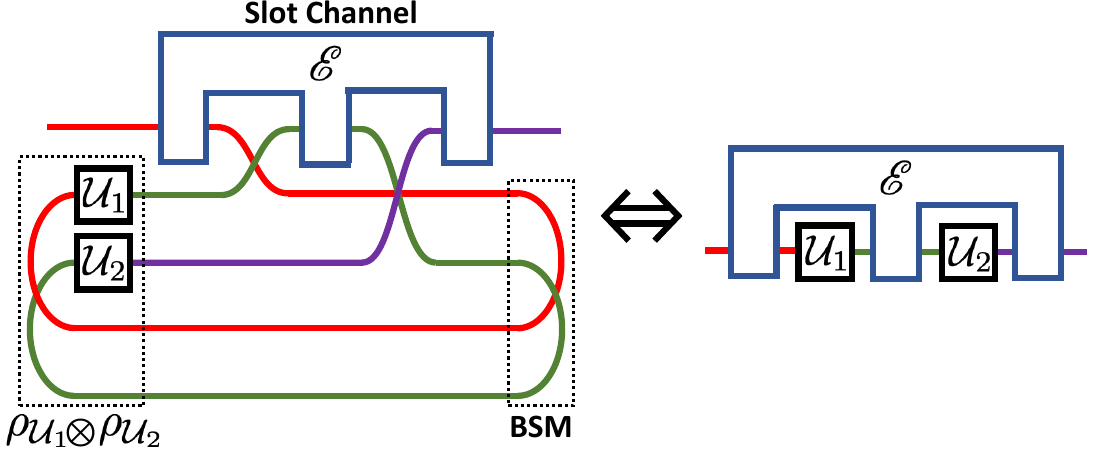}
\caption{The action of a quantum comb represented with slot channel.}
\label{fig:slot_iso}
\end{figure}

From now on, to ensure the clarity of our expression, we will use the process matrix notations, where operators $\rho$ are written as ket vectors $\pket{\rho}$. Suppose we want to apply a $M$-comb non-Markovian noise to $M-1$ different ideal circuit layers $\mathcal{U}_1, ..., \mathcal{U}_{M-1}$. The $m$-th circuit layer can be written in terms of its unnormalised Choi state that lives in the $(m+1)$-th and $(m+M)$-th registers
\begin{align}
    \pket{\rho_{\mathcal{U}_m}} = (\mathcal{U}_{m}^{(m+1)} \otimes \mathcal{I}^{(m+M)})\pket{\Phi_+^{(m+1,m+M)}}.
\end{align}
Here $\pket{\Phi_+^{(m+1,m+M)}}$ is the unnormalised maximally entangled state as defined in the main text. 
Therefore, we can represent the output channel of the quantum comb as
\begin{equation}
    \begin{split}
        \mathcal{U}_{\varepsilon, 1} 
        &= \left(\bigotimes_{m=2}^{M}\pbra{\Phi_+^{(m,m+M-1)}}\right) \mathcal{C}_{\mathscr{E}}^{S,  (1,\cdots M)} \left(\bigotimes_{m=1}^{M-1} \pket{\rho_{\mathcal{U}_m}}\right)\\
    &= \left(\bigotimes_{m=2}^{M}\pbra{\Phi_+^{(m,m+M-1)}}\right) \mathcal{P}_M^{-1} \mathcal{C}_{\mathscr{E}}^{(1,\cdots M)}\left(\bigotimes_{m=1}^{M-1} \pket{\rho_{\mathcal{U}_m}}\right),
    \end{split}
\end{equation}
which is graphically represented in~\cref{fig:slot_iso}.
Here we see that the additional swaps needed in \cref{eq:choi_to_super} are simply for switching between the Choi channel representation and the slot channel representation. 

We can further define the Choi state $\rho_{\mathscr{E}}$ of a comb $\mathscr{E}$ being the Choi state of its Choi channel $\boldsymbol{C}_{\mathscr{E}}$
\begin{align}
    \pket{\rho_{\mathscr{E}}} = (\mathcal{C}_{\mathscr{E}}^{(1,...M)} \otimes \mathcal{I}^{(M+1,...2M)}) \left(\bigotimes_{m=1}^{M}\pket{\Phi_+^{(m,m+M)}} \right).
\end{align}
Using this Choi state representation, we can look at the whole story in another picture. The ideal quantum circuit can actually be more generally represented as a quantum comb $\mathscr{U}$ with $M$ slots (and thus $M+1$ teeth) with each slot being an error location that can be affected by one of the tooth of the non-Markovian noise. Hence, studying the effect of the non-Markovian noise on the circuit is simply fitting the $m$-th teeth of the non-Markovian noise into the $m$-th slot of the circuit. The resultant noisy circuit can be obtained by applying the slot channel of the circuit to the Choi state of the non-Markovian noise and then perform bell measurement:
\begin{align}
    \mathcal{U}_{\varepsilon, 1} = \left(\bigotimes_{m=2}^{M+1}\pbra{\Phi_+^{(m,m+M)}}\right)\boldsymbol{C}_{\mathscr{U}}^{S,(1,...M+1)}\pket{\rho_{\mathscr{E}}}^{(2,...M+1)}.
\end{align}
We can also obtain an equivalent result by applying the slot channel of non-Markovian noise onto the Choi state of the circuit. In this picture of applying a slot channel onto a Choi state, if the slot channel has more teeth than the Choi state, then we will have a comb (which includes channels) as the output after performing the bell state measurement. Otherwise, we will have the equivalent Choi state as the output instead. 

In the non-Markovian quantum dynamics literature, the central object to study is the Choi operator of the dynamics $\mathscr{E}$. It is essentially equivalent to the Choi state we defined by switching over the reference qubits and the qubits that the Choi channel acts on.

\end{document}